\documentclass[a4paper,twocolumn,11pt,accepted=2023-07-12]{quantumarticle}
\pdfoutput=1
\usepackage[utf8]{inputenc}
\usepackage[english]{babel}
\usepackage[T1]{fontenc}
\usepackage{amsmath}
\usepackage{hyperref}

\usepackage{tikz}
\usepackage{lipsum}
\usepackage[numbers,sort&compress]{natbib}

\usepackage{
    physics,
    graphicx,
    multirow,
    tabularx,
    mathtools, 
    placeins, 
    nicefrac, 
    hyperref, 
    threeparttable, 
    siunitx,
    enumitem,
    comment,
    amsfonts
}

\usepackage[capitalize]{cleveref}
\Crefname{section}{Sec.}{Secs.}

\hypersetup{
    colorlinks,
    linkcolor={blue!50!black},
    citecolor={blue!50!black},
    urlcolor={blue!80!black}
}

\usepackage[caption=false]{subfig} 
\graphicspath{{Figures/}}

\begin{document}


\title{Adaptive surface code for quantum error correction in the presence of temporary or permanent defects}

\newcommand{\oxddress}{\affiliation{Department of Materials, University of Oxford, Parks Road, Oxford OX1 3PH, United Kingdom}}
\newcommand{\qmaddress}{\affiliation{Quantum Motion, 9 Sterling Way, London N7 9HJ, United Kingdom}}

\author{Adam Siegel}
\email{adam.siegel@materials.ox.ac.uk}
\oxddress
\qmaddress

\author{Armands Strikis}
\email{armands.strikis@materials.ox.ac.uk}
\oxddress

\author{Thomas Flatters}
\email{thomas.flatters@materials.ox.ac.uk}
\oxddress

\author{Simon Benjamin}
\email{simon.benjamin@materials.ox.ac.uk}
\oxddress
\qmaddress

\date{July 15, 2023}

\begin{abstract}
    Whether it is at the fabrication stage or during the course of the quantum computation, \textit{e.g.} because of high-energy events like cosmic rays, the qubits constituting an error correcting code may be rendered inoperable. Such defects may correspond to individual qubits or to clusters and could potentially disrupt the code sufficiently to generate logical errors.
    In this paper, we explore a novel \textit{adaptive} approach for surface code quantum error correction on a defective lattice. We show that combining an appropriate defect detection algorithm and a quarantine of the identified zone allows one to preserve the advantage of quantum error correction at finite code sizes, at the cost of a qubit overhead that scales with the size of the defect. Our numerics indicate that the code's threshold need not be significantly affected; for example, for a certain scenario where small defects repeatedly arise in each logical qubit at a relatively high rate, the noise threshold is $2.7\%$ (versus the defect-free case of $2.9\%$). We also confirm a strong sub-threshold scaling, with a code distance reduction of the order of the defect size only. These results pave the way to the experimental implementation of large-scale quantum computers where defects will be inevitable.
\end{abstract}

\maketitle

\section{Introduction}

Quantum computers must become fault tolerant in order to be stable enough to run deep quantum algorithms. Logical gate error rates of $10^{-15}$ or lower are then feasible, allowing the execution of quantum algorithms such as Shor's and Grover's \cite{Fowler_2012, resource_estimation_shor_algo, google_surface_code}. This is made possible by assembling a considerable number of qubits together to form error correcting codes \cite{shor_QEC, Terhal_2015, Campbell_2017}. In these large structures, it may be unrealistic to assume that all components will be working nominally after their fabrication. Even if this were the case, some events may temporarily or permanently disrupt their normal behaviour in the course of the computation: high-energy events leading to a surge of correlated errors, like cosmic rays in superconducting and silicon devices \cite{Martinis_2020, Wilen_2021, McEwen_2021}; or leakage and loss of qubits, for instance in ion traps or neutral atom arrays \cite{Vala_2005,Bermudez_2017,Cong_2021, Brown_2019}. In both cases, these events create \textit{defects} on the lattice, either preexisting the computation or occurring while it is running. These defects must be distinguished from ones that can voluntarily be introduced in the surface code as a means to store logical information \cite{Wooton_2017}. As experimentally identified in the repetition code \cite{acharya2022suppressing}, the uncontrolled defects studied in this paper can alter the code's performance so profoundly that it eliminates the exponential suppression of errors upon which deep computations rely. For any hardware platform where this occurs, some protocol must be designed to deal with these defects.

The case of `chip-level' errors was studied in \cite{Xu_2022}, where it was shown that concatenating multiple codes over separate chips would effectively reduce the rate of catastrophic events. Our paper is complementary as it studies smaller defects, that do not disrupt an entire chip. In this scenario, the goal is to retain the error correction capability in a code where some stabilisers cannot be measured or would just yield essentially random results. In both cases, our approach is to disregard all these \textit{faulty qubits} manifesting the defect and remove them from the code.

This picture has already been investigated for the surface code in \cite{Stace_2009, Nagayama_2017, Auger_2017}. These papers show that even when some defective zones of fixed size are removed from the code, effectively creating punctures in it, the code remains robust to errors as long as a modified set of stabilisers is measured. These punctures do have the effect of lowering the distance of the code. However, it remained unclear if defects of fixed density --- rather than size --- would disrupt the exponential suppression of errors of the surface code in the asymptotic limit. Recent work~\cite{Strikis_2021} introduced a method based on code deformation involving shells that isolate defects. That paper provided an analytic proof that a threshold must exist using their methodology. However, the proof involves a series of compounding worst-case assumptions so that a realistic estimate of the threshold could not be obtained. Moreover, obtaining a meaningful threshold requires the creation of bespoke decoders; in doing so there is the opportunity to generalise from factory defects to encompass defects happening on the fly.

In the present paper, we tackle these questions and establish that a realistic high defect rate threshold does indeed exist. We numerically compare both of the previous approaches and show in which regime one is outperformed by the other, focusing on the case of memory storage --- we do not study the resilience to errors when quantum gates are implemented. This knowledge, together with a novel defect detection algorithm that recognises events occurring during the execution of the stabiliser measurements, enables us to design an \textit{adaptive surface code} that deforms whenever a defect is detected so as to exclude it from the code. We then exhibit a threshold for defects detected during these stabiliser measurements and compare it to that of a defect-free surface code. On top of this, we estimate the resource overhead needed to overcome these defects. Ultimately, we were able to numerically exhibit the existence of a \textit{practical} threshold, i.e., one that would not impede the use of error correcting codes in the presence of defects, which, again, will be inevitable.

\section{Methods}

In this section, we present the general workflow of our adaptive surface code. 
It can be split in three distinct steps: first, the detection of the defective zone (if any); second, a code deformation allowing one to exclude the identified defect and thus store the logical information in the remaining qubits; third, the computation of the relevant stabiliser events and the deduction of a correction.

\subsection{Formalism} \label{section: formalism}

First, we clarify the formalism that will be used in the rest of this section. In the surface code, the logical information is stored in data qubits and is controlled by a group of commuting operators called \textit{stabilisers}. This group is generated by 4- (or 2- on the edge) body operators that measure the parity of the data qubits in the $Z$ or $X$ basis. In Figs. \ref{defect_detection} and \ref{code_deformation}, the data qubits are located on vertices and the stabiliser generators measure the parity of all data qubits around a given face. The code space, that is the common +1 eigenspace of all stabilisers, has dimension 2 and is spanned by the logical $X$ and $Z$ operators. They are stringlike operators that commute with all stabilisers but anticommute with each other (green and red lines in Fig. \ref{code_deformation}). The minimum length of any non-trivial logical operator is called the code distance.

If qubits are disabled to avoid a surge of errors in case of defect, the number of degrees of freedom in the code space increases, effectively adding \textit{gauge operators} to the code: these operators still commute with all the stabilisers but can anticommute with each other. They are represented in brighter colours on the left-hand side of Fig. \ref{code_deformation}. Interestingly, \cite{Stace_2009, Nagayama_2017, Auger_2017} noted that, when puncturing a hole in a surface code, the product of the gauge operators around the hole however commutes with all the stabilisers and gauge operators, hence being itself a stabiliser of the code, called \textit{superstabiliser} (see Figure \ref{code_deformation}). In practice, the measurement of the superstabiliser is performed by measuring the individual gauge operators around the hole, and computing the product of these outcomes. As $X$- and $Z$-basis operators do not commute (only their product does), they have to be evaluated at different times. Measuring one subset of commuting gauge operators is called \textit{gauge fixing} as --- without errors --- it fixes the otherwise random value of these operators when repeating their measurement.

\subsection{Defect detection}

\begin{figure}
    \centering
    \includegraphics[width=\linewidth]{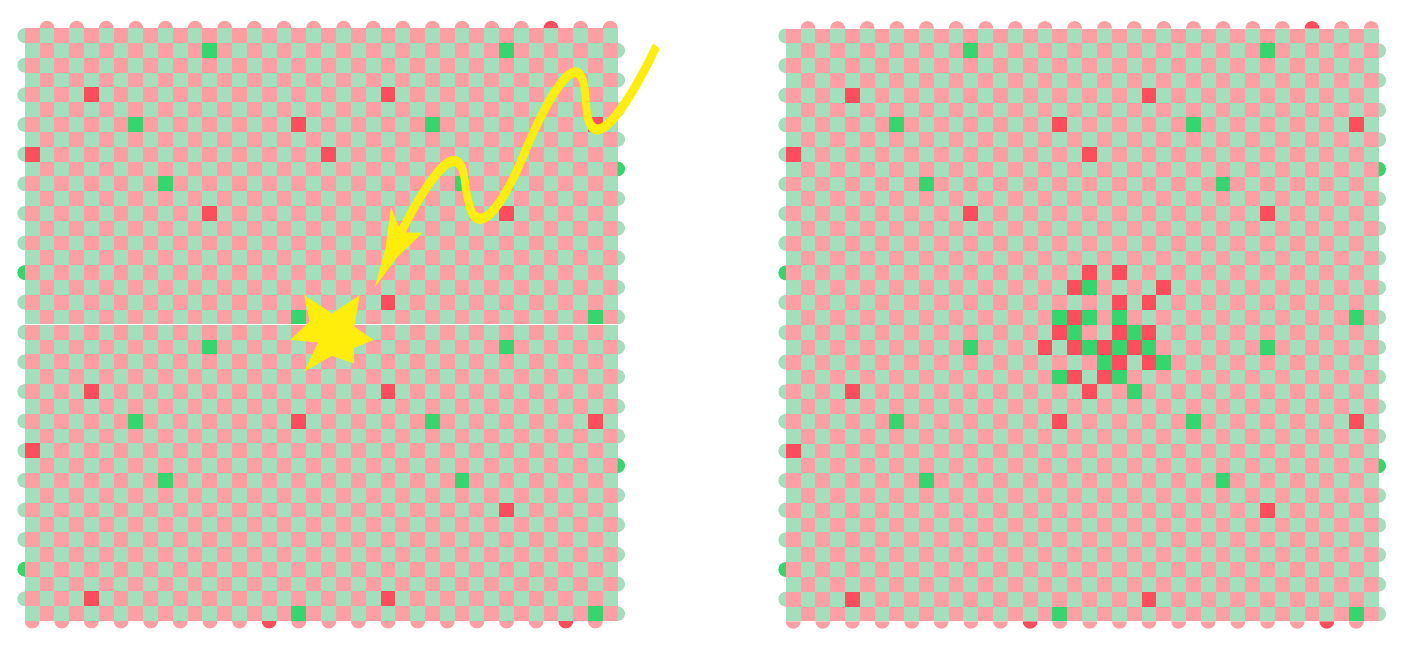}
    \caption{Rotated surface code before and after an abnormal event. The stabilisers are located on faces and a brighter color indicates a non-trivial syndrome measurement.}
    \label{defect_detection}
\end{figure}

Two different types of defects are distinguished in this paper: defects that happen \textit{before} the code is run and that are detected offline (\textit{e.g.} fabrication defects), and defects that happen \textit{while} the code is running and that must be identified on the fly (\textit{e.g.} cosmic rays).
We emphasise that by `defect' we refer to a cluster of qubits that are rendered effectively inoperable, either permanently or for a finite time; this is in contrast to the finite rate of transient errors which is presumed to afflict all qubits during the normal operation of the machine. In practice, it will be relatively straightforward to detect pre-existing defects: post-fabrication analysis by the manufacturer (quality control), and system calibration prior to a computation. The greater challenge is the real-time detection while the code is running. The approach obviously depends on the nature of the defect. We focus on a case known to occur: an instability in qubit(s) corresponding to noise far above the normal level, \textit{i.e.} the behavior seen after a cosmic ray impact.
In order to keep our method as general as possible, it is assumed that the only available  information is the stabiliser measurements. Defects in more specific systems could however be detected more efficiently with tailored protocols \cite{McEwen_2021}. A defect will be modelled via the appearance of a large density of correlated syndrome events in spacetime at a given location of the code (see Figure \ref{defect_detection}).
The challenge is then to identify all the faulty qubits but to avoid misidentifying normal qubits as faulty --- and to do so in minimal time if the detection algorithm is run during the computation (rather than, say, in a calibration phase). To achieve this, stabilisers experiencing more than $n_\text{flips}$ flips within a time $\Delta t_\text{flips}$ are first identified, then gathered in clusters via the DBSCAN algorithm (density-based spatial clustering of applications with noise). 
This allows us to group the previously identified stabilisers in clusters of defects (in cases where multiple defects have occurred) while removing any isolated stabiliser that could have flipped above $n_\text{flips}$ but without being a part of the defective zone. All the hyper-parameters of the detection process ($n_\text{flips}$, $\Delta t_\text{flips}$, and the parameters of the DBSCAN algorithm) are chosen offline by the user to maximise the detection performance. For example, typically in our experiments we set $n_\text{flips}=3$ and $\Delta t_\text{flips}=6$, meaning a stabiliser is considered faulty if it flipped more than 3 times within 6 consecutive rounds.

\subsection{Code deformation} \label{code_def}

After the defect is detected, the second step consists of excluding the faulty data and ancilla qubits from the surface code. This is done by code deformation and two different pictures must be distinguished: defects located inside the code and defects touching the boundary of the code. For simplicity, we will consider the case of defects bounded by square regions in this paper, although the concepts generalise to any shape.

\paragraph{Defects located inside the code}

This case was first reviewed in \cite{Stace_2009, Nagayama_2017, Auger_2017} and uses the concept of gauge operators \cite{Paetznick_2013,Higgott_2021} and superstabiliser detailed in Section \ref{section: formalism}.
When puncturing a hole in a surface code, operators around the hole cannot be used as stabilisers as they do not commute with each other anymore. Rather, a new stabiliser can be evaluated to avoid errors strings terminating at the hole without being detected: the superstabiliser. Its value is inferred from the individual value of each gauge operator. However, as $X$- and $Z$-basis operators do not commute (only their product does), they have to be evaluated at different times.
In this first approach, the two bases are measured on alternating rounds.  Consequently there is no context allowing one to validate the individual measurements. Consider the measurement of a given $X$-basis operator; immediately after, the qubits involved will  be measured according to $Z$-basis operator(s). When our $X$-basis operator is next measured, its value will be independent of its prior outcome due to the non-commutation of the operators -- thus a faulty measurement has no signature or `evidence' on the measurement outcome of an individual gauge operator. Instead, a faulty measurement changes the superstabiliser outcome that the gauge operator belongs to. In consequence, the inferred value of each superstabiliser becomes less reliable as the size of the puncture is increased, since more and more potentially-faulty measurements are involved in the inference of the superstabiliser's value. Nevertheless, the approach is attractively straightforward. We will call this the \textit{basic approach}.

In order to mitigate the increase of faulty superstabiliser measurements, it was recently proposed \cite{Strikis_2021} to keep measuring the gauge operators for a number of consecutive rounds that scales with the size of the puncture, effectively creating alternating blocks of repeated measurements of $X$ (and then, $Z$) gauge operators in spacetime, called \textit{shells}. In the absence of errors, the successive gauge operator measurements in the same basis will yield the same result; thus, errors on gauge operators become easily detectable simply by repeating and comparing the operator measurements. We call this the \textit{shell approach}. It was proved to outperform the basic approach in the asymptotic limit, but finite code sizes were not studied. We report the conclusions of our numerical modelling of such scenarios in the Results section.

\paragraph{Defects located at the boundary}

\begin{figure}
    \centering
    \includegraphics{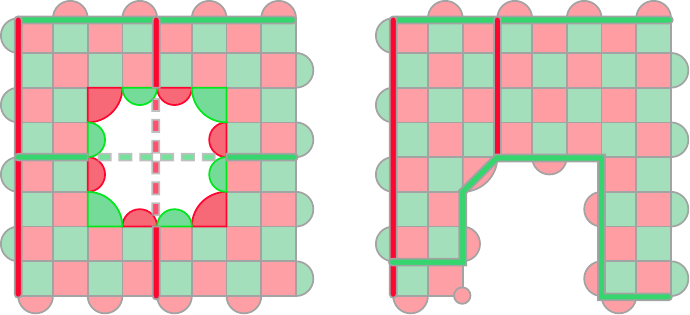}
    \caption{Surface code after defect identification and code deformation. Two cases are represented, depending on the defect's location with respect to the code. $X$ (resp. $Z$) stabilisers and gauge operators are represented in red (resp. green). The gauge operators forming a superstabiliser are in brighter colors (left panel). Some logical operators are drawn with horizontal and vertical lines. Two of them are split in two halves connected by a dashed line. Note that in each case the code distance is $d=L-l$ where $L$ and $l$ are respectively the sizes of the code and the defect.}
    \label{code_deformation}
\end{figure}

The previous ideas do not apply to defects touching the boundary. Let us call \textit{X-boundary} a boundary where $X$ error strings can terminate without being detected (top and bottom boundary in Figure \ref{code_deformation}) -- and similarly for a \textit{Z-boundary}. Along an $X$- (resp. $Z$-) boundary, the product of the $Z$ (resp. $X$) gauge operators does not commute with some of the $X$ (resp. $Z$) gauge operators.
Hence, they cannot form a superstabiliser and need not be measured. After the deformation, the code just looks like a normal surface code with a deformed boundary (see Figure \ref{code_deformation}). Along an $X$-boundary, the $Z$ gauge operators are not measured, which means that the $X$ stabilisers of the initial code do not lose their commutation properties. This is why the distance associated with the logical $Z$ operator is not lowered when an $X$-boundary is deformed. The same applies for the $X$ distance and a $Z$-boundary. Note that when a defect hits a corner of the code, one can choose which boundary to deform.

\subsection{Syndrome calculation and decoding}

Once the code deformation has been determined, the quantum computation can proceed fault-tolerantly. The ongoing process consists of gathering a suitable set of stabiliser and gauge operator measurements after a period of computation, finding the relevant syndrome and decoding it. In the numerical modelling we presently describe, we simply decode using the entire syndrome (since we do not simulate codes so large, nor sequences of stabiliser cycles so long, that this is challenging for the decoder). Regarding the stabiliser and gauge operator measurements, they are obtained in the usual way as endpoints of error strings --- excluding the superstabiliser measurements for now. The only subtlety is that, at the time of a gauge change, the measurement outcomes of the gauge operators are randomised (recall that only the value of the superstabiliser is constant), because of their non-commutation. A difference syndrome can then be computed and used as an input of the decoder.

\begin{figure}
    \centering
    \includegraphics[width=\linewidth]{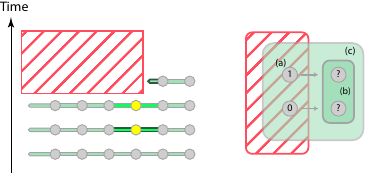}
    \caption{First panel: data qubits are represented with grey circles and stabilisers with green lines. The yellow dots mean an error happened on a data qubit inside the puncture (hatched zone) near its edge and before it starts. If the difference syndrome is directly computed, three events are triggered (highlighted in dark green). As this number is odd, MWPM will have no choice but to link one of them to the boundary, drastically increasing the chance of a logical error. Second panel: the data qubits inside the puncture (a) are measured out in the relevant basis. Their parity can be multiplied with the parity of the gauge operator measured after the code deformation (b) to infer the value of the original total stabiliser (c).
    }
    \label{syndrome}
\end{figure}

This approach can readily be applied if the code deformation was implemented before the stabiliser measurements started (to isolate fabrication errors, say), but not if the defect was detected while they are being measured. Indeed, it is not obvious how to compute the difference syndrome for gauge operators whose number of qubits changed during the code deformation. \cite{Vuillot_2019} proposes a general decoder for gauge changes but it supposes that the whole round of stabiliser measurements before the code deformation is perfect. This allows one to decode the syndrome before the gauge change, then change gauges, then keep the computation going in the new gauge and decode again at the end. This hypothesis is however not suitable to our study as the code deformation happens exactly when the highest amount of correlated errors is occurring. To circumvent this issue, one can note that, as the data qubits inside the defect are to be discarded after the code deformation, they can be measured out in the $(\ket{0},\ket{1})$ or $(\ket{+},\ket{-})$ basis. By doing so one can infer the parity of the original stabiliser from the measurement of the lower-weight gauge operator and the parity of the measured data qubits (see Figure \ref{syndrome}). This value is the appropriate one for the computation of the difference syndrome.

This syndrome can then be input to any standard decoder. In this paper, we used Minimum Weight Perfect Matching (MWPM) \cite{Fowler_2012}. The weights in the matching graph are given by the shortest distance between two points in spacetime. In the basic approach, the superstabiliser replaces the gauge operators in the matching graph as no information is gained from their individual values since they are switched at each round. In the shell approach however, the gauge is fixed for some time hence it is important to include each gauge operator. The only subtlety is that weight-zero matching must be possible each time a gauge is switched. This way, gauge operators taking random values after a gauge change can be linked with each other with no cost. Hence, in the shell approach, the superstabiliser value is never computed; it is instead always broken down into individual gauge operators. In this, our work differs from detector-based decoding \cite{Higgott_2021, mcewen2023relaxing} where no randomisation nor weight-zero edges are needed, as superstabilisers replace gauge operators in the syndrome at the time of a gauge change. Detectors enable faster decoding as less data is fed into the decoding algorithm, but our method is attractively straightforward to implement, as the vertices of the matching graph need not be changed when a defect hits the code or during any of the subsequent gauge changes. Only the edges need to be updated, to account for the new distances between stabilisers or gauge operators when the defective components are turned off, and to include the weight-zero edges. In terms of the output, both approaches are however equivalent as weight-zero edges will prioritise a matching between gauge operators, as if they formed a single vertex in the matching graph.

Finally, when detecting defects in real-time, numerous errors may typically occur between the time a defect strikes the code and when the code is deformed. As a natural improvement to the decoding routine, a weight-zero zone can be added at the location of the defect for a time interval that is needed to detect a defect. This means that all detection events involving qubits, that were \textit{a posteriori} deemed as faulty and triggered a higher number of errors, will be more likely to be linked by MWPM. 

\section{Results}

To verify the performance of our method, we proceeded in several steps. First, we compared the basic and the shell approaches (as defined earlier) and explored which approach had better performance at a given finite scale. This allowed us to decide which one should be used in our adaptive method. With this knowledge, we were then indeed able to implement the adaptive method; as previously explained, we distinguish two types of defects: permanent defects identified before the code is run, solved by a so-called \textit{static approach}, and defects happening while the code is running and that we must detect, tackled by the \textit{adaptive approach}.

In all of the following simulations, we focus on $X$ errors only -- this is possible since the surface code can be regarded as a means to protect against $Z$ and $X$ errors as two separate (yet interlaced) tasks; for any homogeneous error model, the performance with respect to $Z$ errors will be identical to the performance versus $X$ errors. We adopt a simple phenomenological error model where errors are independent and identically distributed, and afflict data and ancilla qubits with the same probability $p$ at each time step. When a defect occurs during the computation, it manifests as an abrupt increase in the phenomenological error rate to $q=0.5$ for the affected qubit(s). For simplicity, we assume these qubits form a square of side $l$ centred at a random location of the code. Not more that one defect will be simulated in this paper due to the increasing complexity of the decoder in the presence of multiple defects. Finally, each data point in the following plots is obtained from Monte-Carlo simulations with a number of runs ranging between 10,000 and 100,000 (depending on the expected logical error rate). Error bars, when plotted, show the variance of the sampled data.

\subsection{Basic vs. shell}

\begin{figure}
    \centering
    \includegraphics[width=\linewidth]{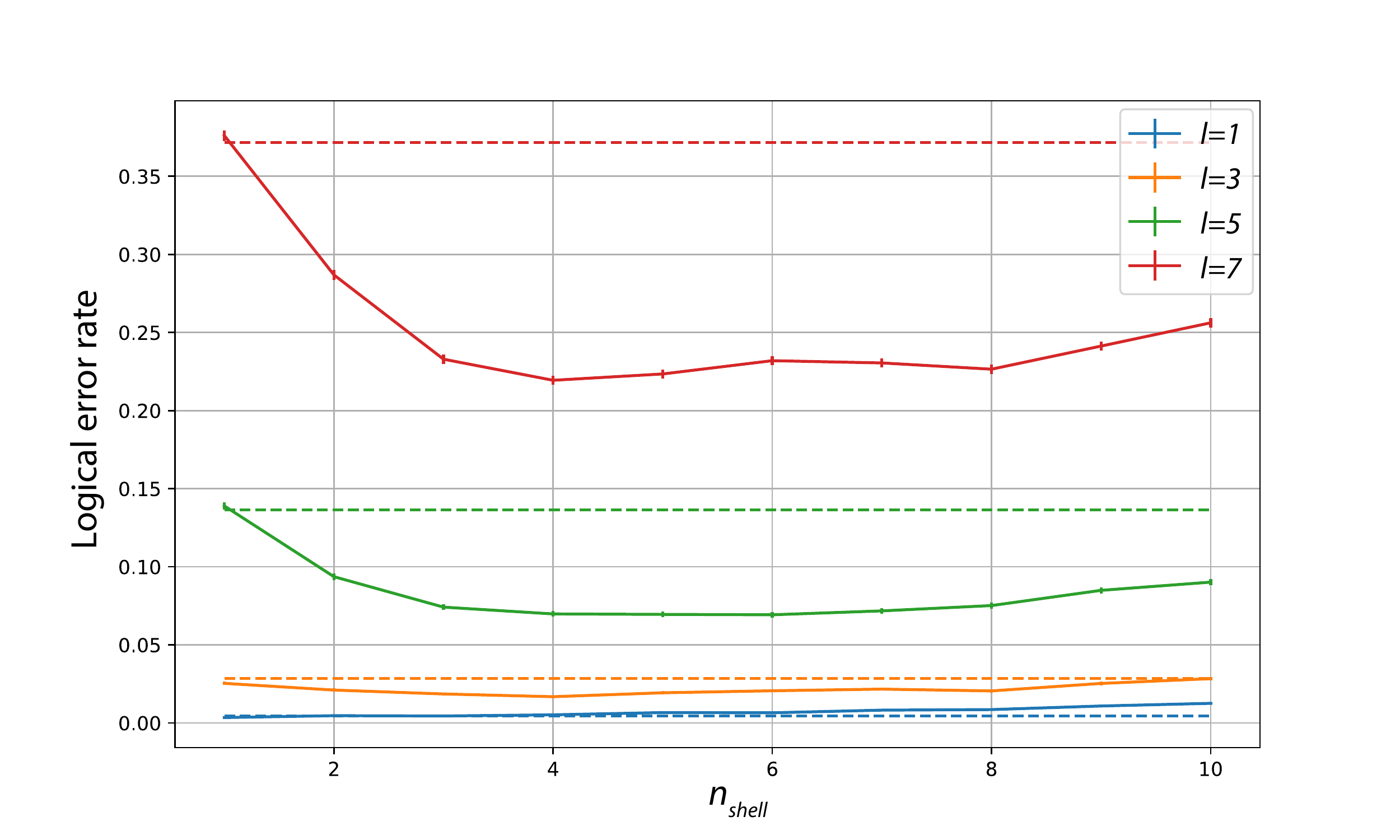}
    \caption{Comparison of the performance of the shell (solid lines) and basic (dashed lines) approaches over the number of measurements per shell $n_\text{shell}$ and for various defect sizes $l$ at $L=15$ and $p=1.5$\%.}
    \label{armands_stace_n_meas_defect}
\end{figure}

An important parameter in the shell approach compared to the basic one is the number of measurements per shell $n_\text{shell}$, \textit{i.e.} the number of rounds for which a type of gauge operator is measured before changing gauges and measuring the other type. Figure \ref{armands_stace_n_meas_defect} shows how the logical error rate evolves with this parameter and for various defect sizes. The logical error of the basic approach is plotted in dashed lines for comparison. We notice that when the size of the defect is increased, the shell method starts to outperform the basic one for the right choice of $n_\text{shell}$. There is an optimal choice of this parameter according to the likely defect size. A too-high rate of gauge switching does not give enough time to infer the value of the superstabiliser accurately. A rate set too low results in, for example, a poor capability to temporally localise the end of a chain of phase flip errors that terminates on the boundary while $Z$-gauge operators are being measured.

\begin{figure}
    \centering
    \includegraphics[width=\linewidth]{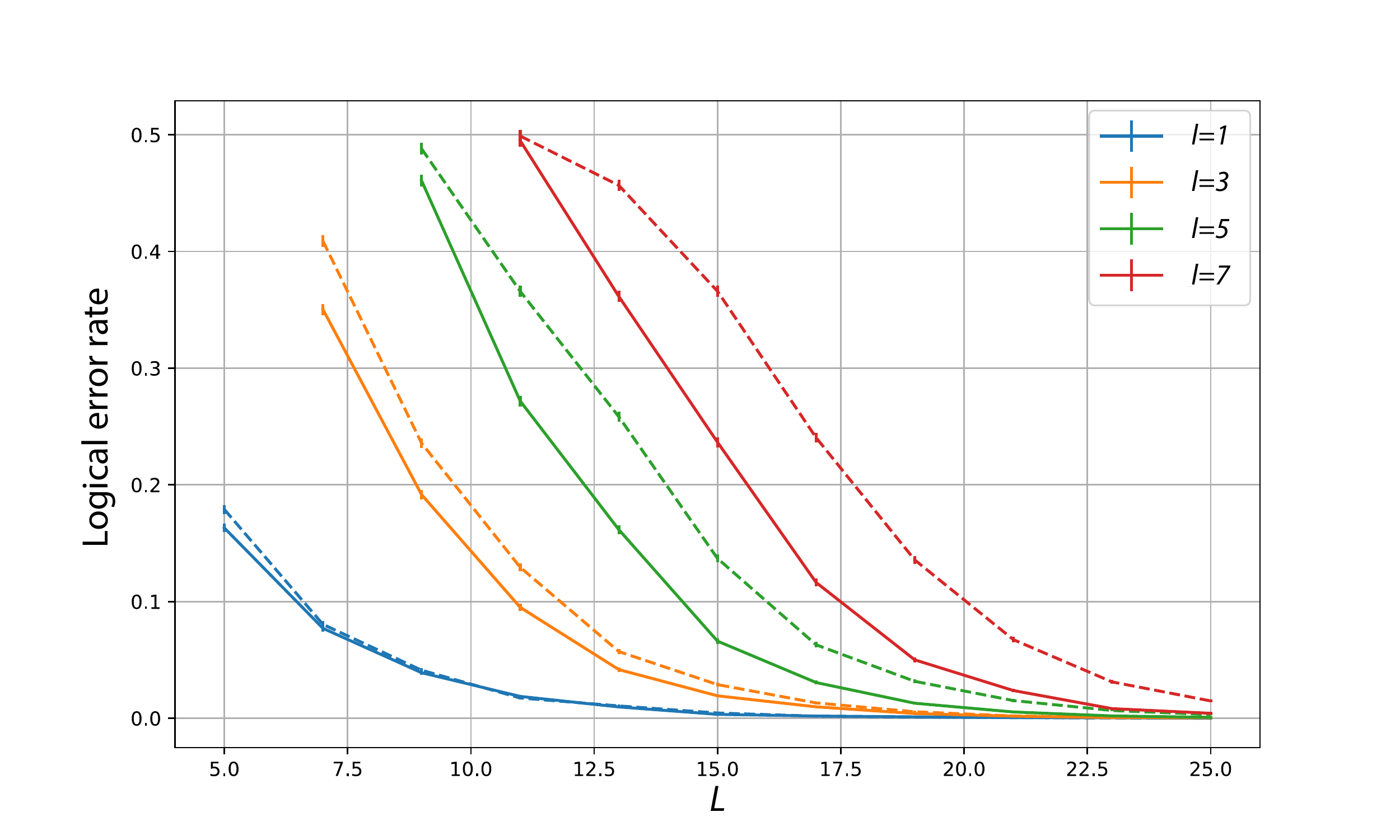}
    \caption{Comparison of the performance of the shell (solid lines) and basic (dashed lines) approaches over the size of the code $L$ and for various defect sizes $l$ at $p=1.5$\%.}
    \label{armands_stace_L}
\end{figure}

The data shown in Figure \ref{armands_stace_n_meas_defect} allows us to infer the proper choice: it is favourable to choose the shell method for defect sizes $l\ge3$ with a number of measurements per shell scaling with $l$. We set $n_\text{shell}=l$ in all the subsequent simulations. For this choice of parameters, we show that the difference between the shell and the basic approaches scales with the defect size, as one would expect (Figure \ref{armands_stace_L}).

\subsection{Qubit overhead for quantum error correction on a defective lattice}

\begin{figure}
    \centering
    \includegraphics[width=\linewidth]{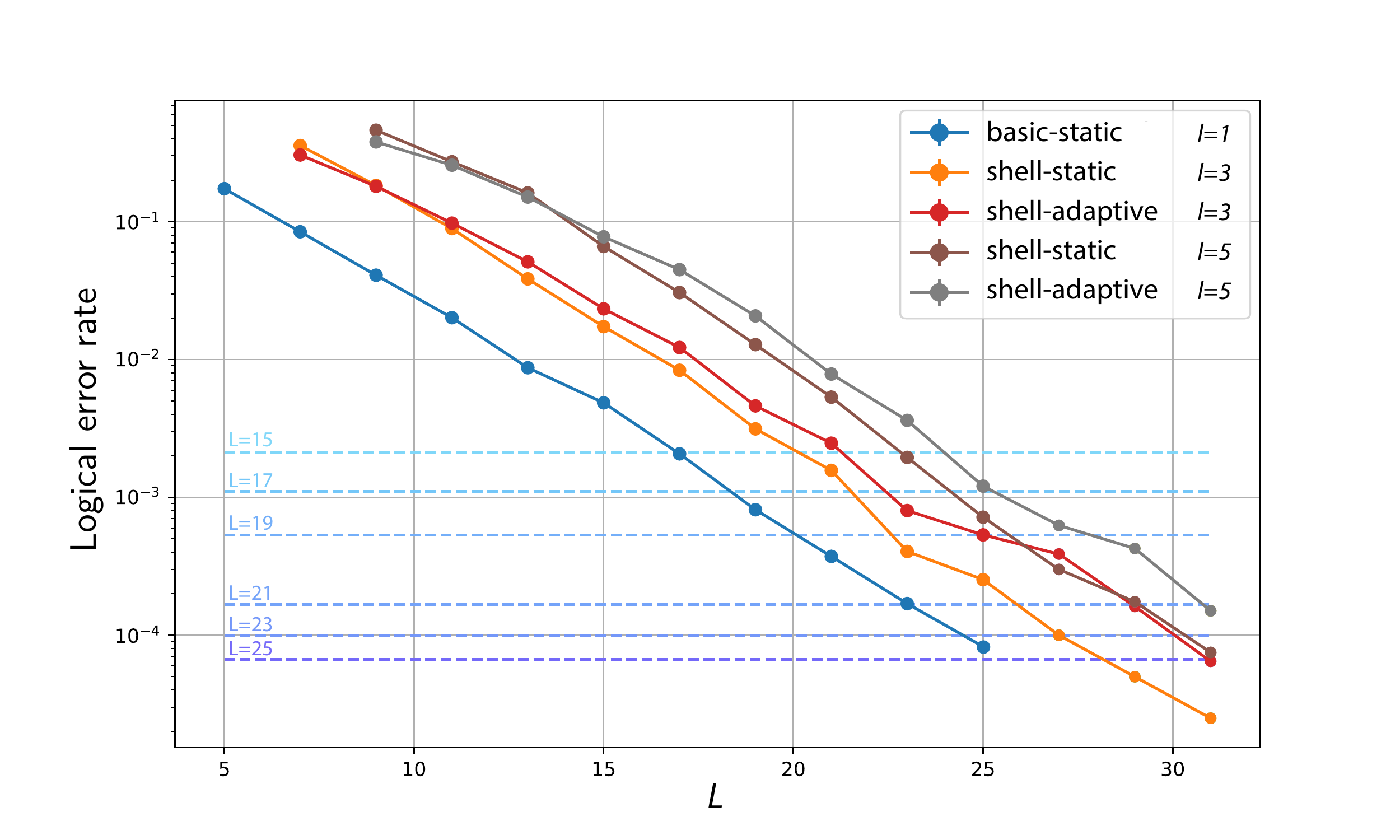}
    \caption{Logical error rate over $L$ for three different defect sizes, $1\times 1$, $3\times 3$ and $5\times 5$, with the high-performing protocols for each scenario selected (\textit{i.e}. simple superstabiliser for the smallest defect, and the static or adaptive shell approaches otherwise). The dashed lines respectively correspond to the logical error rate achieved by a regular surface code of size 15, 17, 19, 21, 23 and 25. In all cases, qubits suffer `normal' phenomenological noise  at 1.5\%. The intersection of a dashed line with a solid line gives us the enlarged code size $L$ needed to replicate the performance of the smaller code on a defect-free device.}
    \label{compare_with_no_hole}
\end{figure}

To compensate for the loss of protective power due to the presence of defects, it will be necessary to increase the number of qubits in the code. However, for our methods to be of practical use, this overhead must be kept as low as possible: it is estimated in Figure \ref{compare_with_no_hole}, where the logical error rates of the different approaches presented in this paper are compared to the logical error rate achieved by different sizes of regular surface code. The best form of mitigation strategy, as determined by the prior results, is applied in each case according to the defect size.

We first observe that the performances of the static and adaptive approaches are very close, showing the efficiency of the latter to detect defects in real time. The divergence for large $L$ stems from imperfect defect detection, specifically setting up larger shells than necessary around defects --- if a qubit far away from the defect happens to `flicker', due to random noise, it may be included within the inferred bounds of the defect, hence creating a significantly larger shell than necessary. Note that this is an artefact of our simulations where at most one shell is employed; moreover the effect could certainly be mitigated by the use of more powerful detection and inference methods that are beyond the scope of this paper.

A second observation is that the adaptive surface code still provides an exponential suppression of errors; this was not an obvious outcome given the existence of a finite period of highly correlated errors between the appearance of the defect and its detection and subsequent code deformation. Third, we note that for any approach, the increase of code size that is necessary to equal the performance of an ideal surface code is of the order of the defect size, which is expected since it is the amount by which the code distance is lowered.

\subsection{Performance of the adaptive surface code} \label{section: threshold}

\begin{figure}
    \centering
    \includegraphics[width=\linewidth]{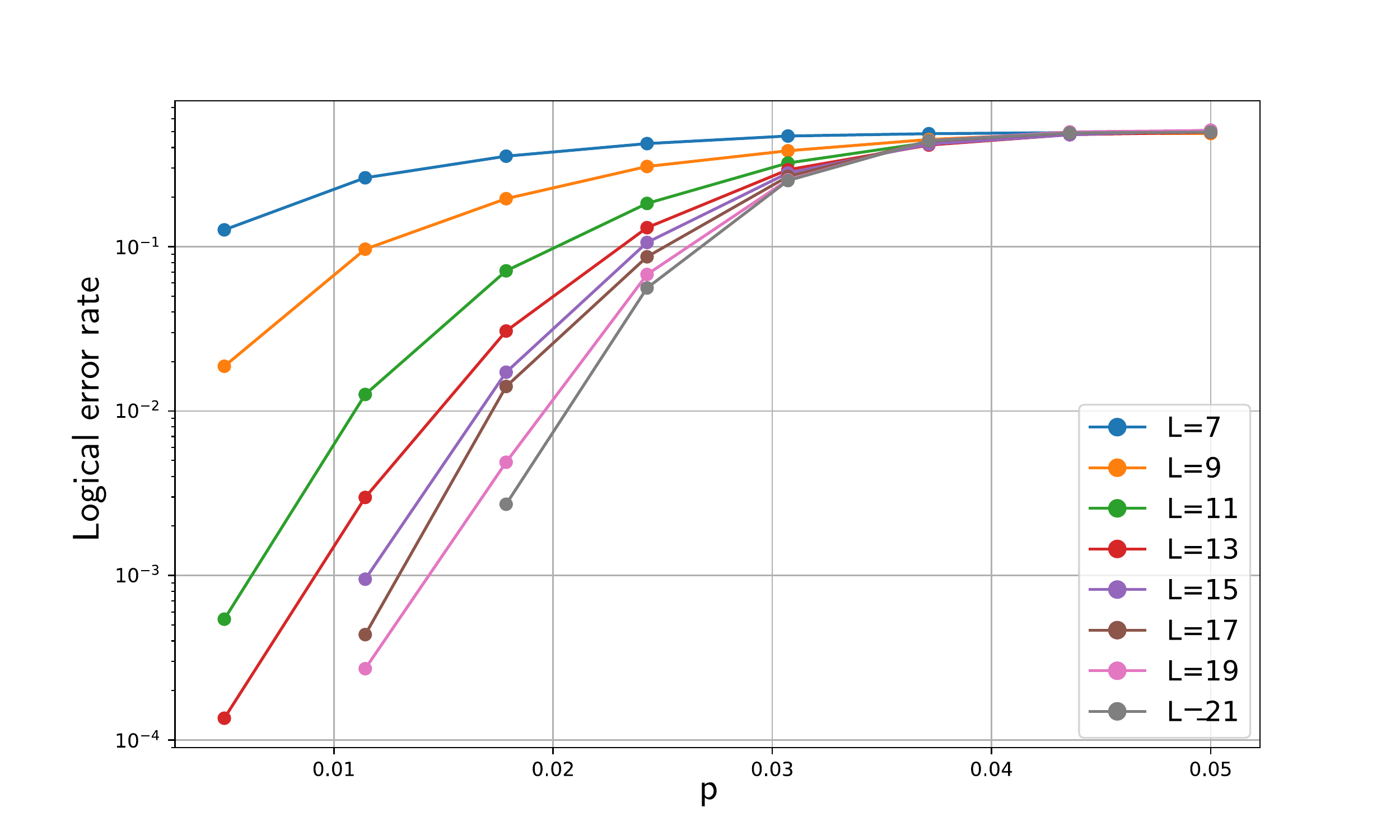}
    \caption{Logical error rate for the adaptive surface code under phenomenological noise for a defect of size $3\times 3$ (ignoring that larger codes are more likely to be defective).}
    \label{cr_th_no_correction}
\end{figure}

The final step is to specifically analyse the performance of the adaptive approach: the capability to detect defects in real time via the DBSCAN algorithm, exclude them from the code, and keep the computation going. The logical error rate is plotted against the physical error rate $p$ in Figure \ref{cr_th_no_correction} for different code sizes $L$. For any value of $p$, large codes appear to perform better than small codes: this is because of the additional burst of errors due to the defect to which small codes are less resilient.

However, in order to examine the existence of a threshold, we must incorporate an additional crucial factor: smaller codes are also less likely to suffer a defect, given their smaller number of qubits. Over the course of a long quantum computation, the code may be hit by some defects, recover from them so that any required code deformation can be reversed, and subsequently suffer further defect events. Without any further assumption on the defects' nature, it is most reasonable to assume they occur uniformly in space and time --- which is certainly the case for specific defects too, \textit{e.g.} cosmic rays. As a result, we can define a constant rate of defects $\rho$ per unit space and time, or equivalently, per qubit and per round of stabilisers. If we also suppose that defects survive in the code for $T$ rounds, then the effective logical error over the whole computation will be given by:

\begin{equation}
    p_\text{log} = \sum_{k\ge0} \langle t_k \rangle \times p(\text{logical error}|n_\text{def}=k)
\end{equation}
\begin{equation}
    \langle t_k \rangle = \frac{\mathrm{e}^{-2L^2\rho T}(2L^2\rho T)^k}{k!}
\end{equation}
with $\langle t_k \rangle$ the proportion of the time spent suffering $k$ defects and $n_\text{def}$ the number of defects in the code. This simple probability model assumes that defects are uniformly distributed in time and mutually independent. In particular, it supposes that they can overlap. See details of Equations 1 and 2 in Appendix \ref{section: prob_multi_defect}.

As we simulate at most one defect in this paper, the conditional probabilities are evaluated with a simple heuristics deduced from the previous subsection: for $k=0$, a simple threshold calculation is performed for a rotated surface code without defects; for $k=1$, the data from Figure \ref{cr_th_no_correction} is used; and for $k\ge2$, we estimate from Figure \ref{compare_with_no_hole} that the code distance is lowered by at most $2l$ in the presence of a defect of size $l$. Focusing on defects of a single size only, we deduce the following rule: the probability of a logical error in the presence of $k$ defects of size $l$ in a surface code of size $L$ is given by the probability of a logical error in the presence of 1 defect of size $l$ in a surface code of size $L-2l(k-1)$. This assumption is more pessimistic than reality as defects overlapping actually involve fewer faulty qubits.

\begin{figure}
    \centering
    \includegraphics[width=\linewidth]{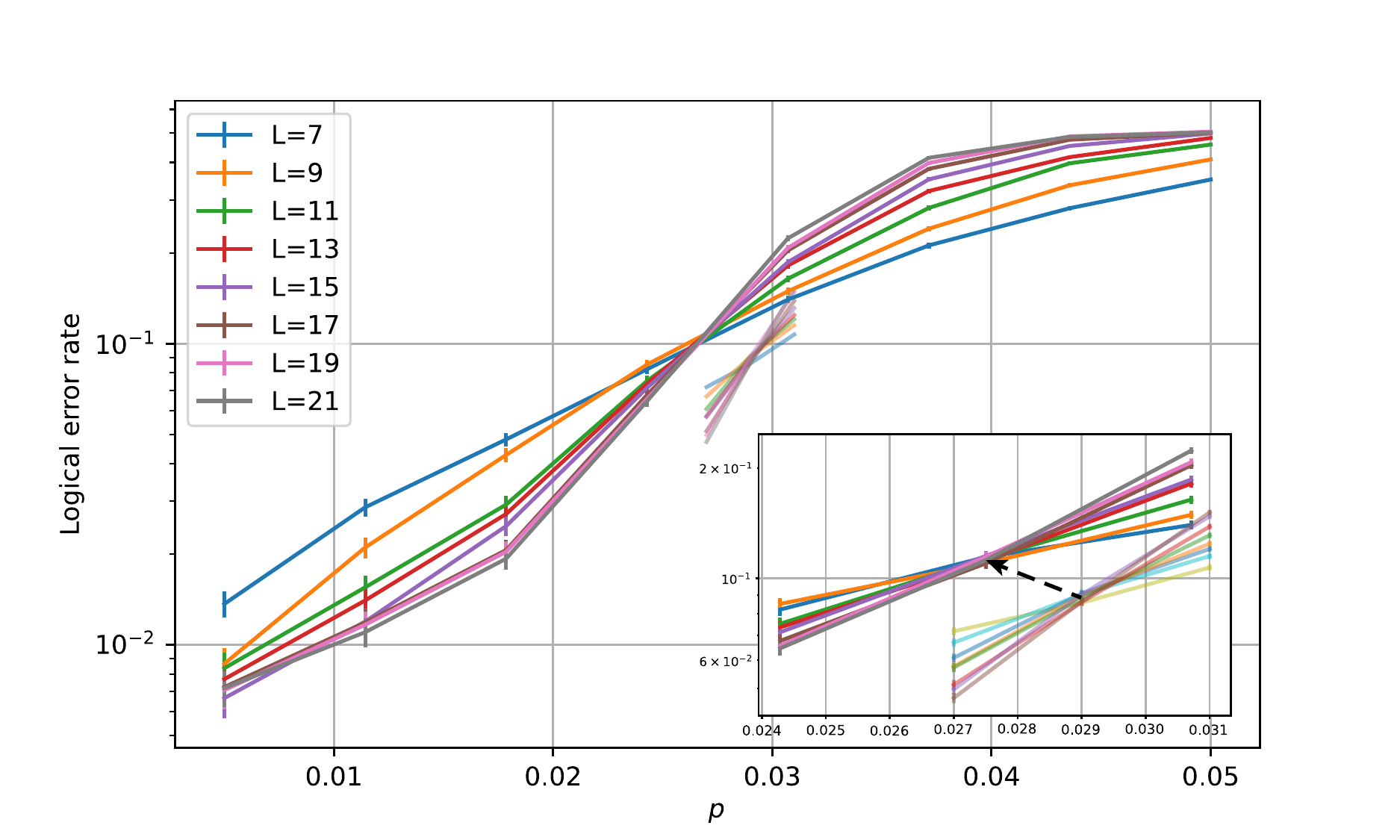}
    \caption{Opaque lines: threshold plots for the adaptive surface code under phenomenological noise for $\rho=10^{-5}$ and $T=100$ (see main text). Faded lines: threshold for the non-defective rotated surface code. Inset: zoom around the threshold region. The dashed arrow represents the evolution of the threshold when the rate of defects is increased.}
    \label{cr_th_corrected}
\end{figure}

Figure \ref{cr_th_corrected} is obtained from this approach and manifests the existence of a threshold. The faded lines correspond to no defect ($\rho=0$) and the opaque lines to a rate of defect nucleation, per qubit and per stabiliser round, of $\rho=10^{-5}$. For logical qubits of size $21\times21$, this is therefore one defect occurring every 110 stabiliser rounds, or around 8800 defects per second afflicting each logical qubit in a quantum computer with a 1MHz stabiliser cycle. The defect survival time $T$ is set to 100, meaning that, each time a defect occurs, it takes 100 stabiliser cycles to reset and reintegrate the afflicted physical qubits. This means spending 37\% of the total time with one defect, 16\% with two defects and 5\% with three: the values of $\rho$ and $T$ are hence not particularly conservative. For this value of $T$, the chosen $\rho$ is also approximately the highest value one can pick for our heuristic model to apply: as logical error rates are evaluated with a relatively pessimistic rule for two defects or more (lowering the code size by twice the size of the defect), scenarios where more than three defects occur must remain quite rare.

In the inset of Fig. \ref{cr_th_corrected}, the black dashed arrow shows the evolution of the threshold when $\rho$ is increased (a more detailed plot can be found in Appendix \ref{section: vary_rho}). It is worth noting that it ranges between $2.7\%$ and $2.9\%$ --- the threshold of the non-defective rotated surface code: in the presence of defects, the threshold is lowered but only by a small amount, which confirms the efficiency of our method, even under fairly pessimistic assumptions.

\bigskip
\section{Discussion}

We have thus established a procedure to deal with defects in the surface code, whether they are present and detected from the start, like fabrication defects, or occurring while the code is running, like cosmic ray impacts. This work is hence relevant to recent experimental studies that identified how damaging this latter kind of defect can be \cite{McEwen_2021, acharya2022suppressing}.
The previous section has shown the performance of our method, in particular with the existence of a threshold that is only slightly lower than that of the non-defective surface code. A resource analysis also showed that the qubit overhead needed to overcome the presence of a defect was moderate, only scaling with the number of defective qubits.

However, our numerical simulations were performed under rather simplistic assumptions: phenomenological noise model for the lattice and the defect, defects are square, only one defect at a time can affect the code. These were only assumed out of simplicity for the numerical implementation, hence we do not think dropping these hypotheses would severely change our results, but it would be valuable to run further simulations relaxing them. In particular, the value of the defect rate $\rho$ was chosen to be the maximum our model could tolerate, by guaranteeing three defects or more remained relatively rare events. Further work could explore higher defect rates by properly implementing a decoder for multiple defects, without having to make use of the heuristic rule established in the previous section.

Finally, the decoding algorithm used in this paper is MWPM. While this is the most canonical choice for studies of the surface code, research is today ongoing to find potentially better decoders \cite{Roffe_2020, Delfosse_2020}, \textit{i.e.} that would run faster, be scalable to very large codes and work as locally as possible to avoid memory and bandwidth issues between the quantum device and the classical processor responsible for decoding \cite{Das_2020, Das_2021, bandwith}. The present work could thus be improved by taking these considerations into account and using an efficient local decoder that would be able to deal with defects. A further interesting question is how this research could be composed with the recently-proposed window decoding \cite{Skoric_2022}.

\bigskip
{\bf Note:} During the latter stages of preparation of this work, the authors became aware of a recent paper~\cite{Suzuki2022cosmic} that also tackles the challenge of cosmic ray impacts at the code level. While the details of that paper's approach differ, it presents a similarly optimistic conclusion regarding the power of detection and adaptive code deformation.

\section*{Acknowledgments}
The authors are grateful to Ben Brown for helpful discussions. They would also like to acknowledge the use of the University of Oxford Advanced Research Computing (ARC) facility \cite{richards_2015_22558} in carrying out this work and specifically the facilities made available from the EPSRC QCS Hub grant (agreement No. EP/T001062/1). The authors also acknowledge support from EPSRC’s Robust and Reliable Quantum Computing (RoaRQ) project (EP/W032635/1) and from the IARPA funded LogiQ project.

\bibliographystyle{unsrtnat}
\bibliography{ref}

\newpage

\appendix

\section{Probability model for the logical error rate in the presence of multiple defects} \label{section: prob_multi_defect}

In Section \ref{section: threshold}, we compute the logical error rate as follows:

\begin{equation}\label{A1}
    p_\text{log} = \sum_{k\ge0} \langle t_k \rangle \times p(\text{logical error}|n_\text{def}=k)
\end{equation}
\begin{equation}\label{A2}
    \langle t_k \rangle = \frac{\mathrm{e}^{-2L^2\rho T}(2L^2\rho T)^k}{k!}
\end{equation}

In Equation \ref{A1}, $\langle t_k \rangle$ is the proportion of the time spent suffering $k$ defects and $p(\text{logical error}|n_\text{def}=k)$ is the probability of a logical error over $L$ rounds of stabiliser measurements for a surface code suffering $k$ defects. Hence, $p_\text{log}$ describes the average probability of a logical error for a surface code potentially hit by multiple defects over $L$ rounds of stabiliser measurements. This is the appropriate quantity to evaluate if compared to more canonical threshold calculations where, in the presence of measurement errors, the surface code is run for a number of rounds proportional to $L$ in order to obtain a neat threshold.

We model $\langle t_k \rangle$ in Equation \ref{A2}. To do so, we suppose that defects are uniformly distributed in time and mutually independent. Realistically, the number of defects $X_i$ hitting the code between stabiliser rounds $i$ and $i+1$ is a random variable following a Poisson law of rate $2L^2\rho$, with $\rho$ the rate of defects per qubit and per round of stabiliser measurements --- as the rotated surface code contains roughly $2L^2$ qubits.
If we assume that each defect survives for $T$ rounds and denote $t_{tot}$ the total time of the simulation, $N_i$ the number of defects at round $i$ and $\mathbb{P}(...)$ a measure of probability, then:

\begin{align}
    \langle t_k \rangle &= \frac{1}{t_{tot}}\sum_i\mathbb{P}(N_i=k)\\ \label{A4}
                        &= \frac{1}{t_{tot}}\sum_i\mathbb{P}\left(\sum_{j=i-T+1}^iX_j=k\right)\\ \label{A5}
                        &= \frac{1}{t_{tot}}\sum_i \frac{\mathrm{e}^{-2L^2\rho T}(2L^2\rho T)^k}{k!}\\
                        &= \frac{\mathrm{e}^{-2L^2\rho T}(2L^2\rho T)^k}{k!}
\end{align}
where, between steps \ref{A4} and \ref{A5}, we use that a sum of $T$ mutually independent Poisson laws of rate $\lambda$ is a Poisson law of rate $\lambda T$.

\section{Evolution of the threshold of the adaptive surface code with the defect rate $\rho$} \label{section: vary_rho}

\begin{figure}
    \centering
    \includegraphics[width=\linewidth]{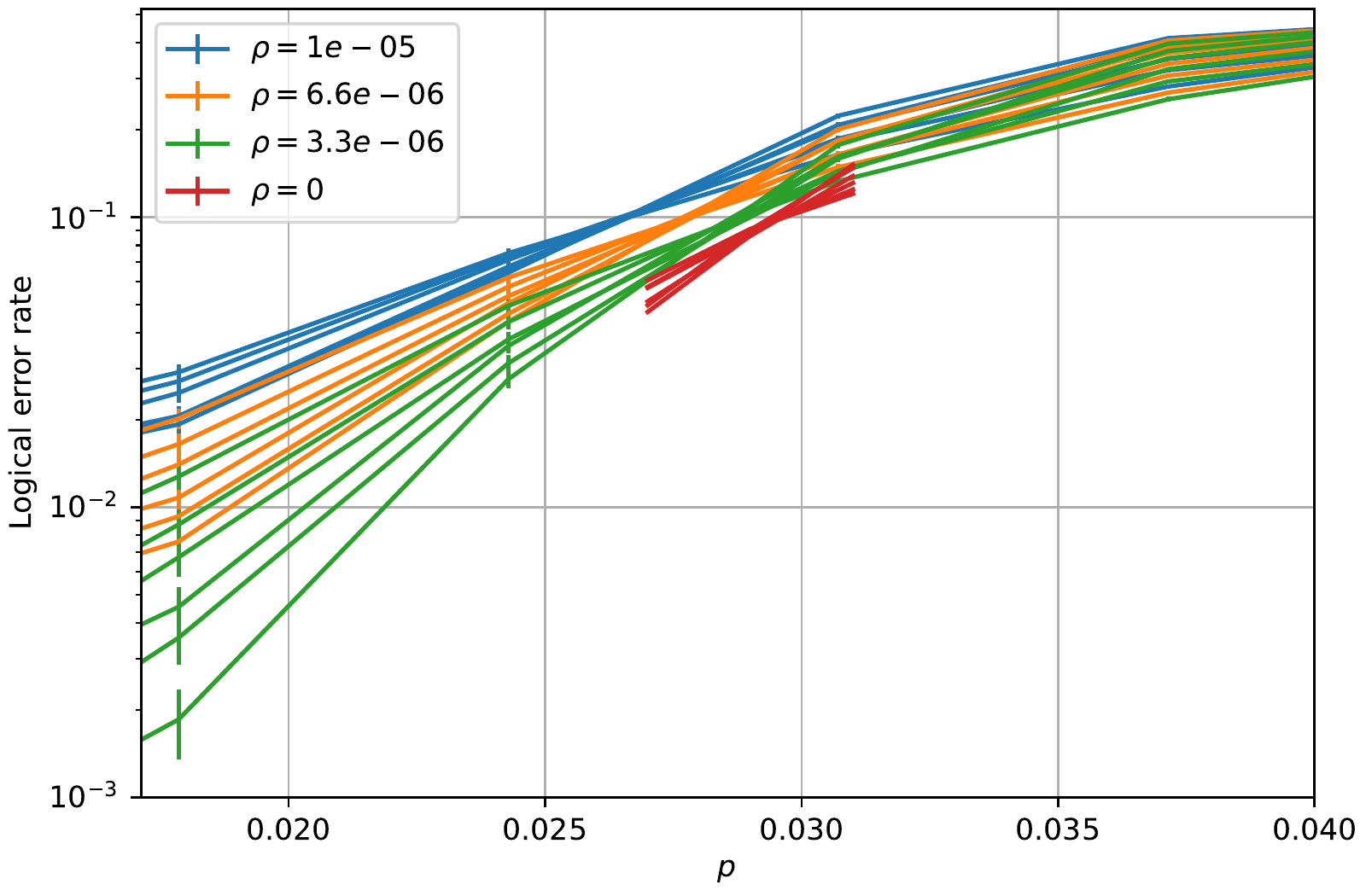}
    \caption{Threshold plots for the adaptive surface code under phenomenological noise for $T=100$ and various values of $\rho$. Lines of the same colour correspond to the same value of $\rho$ but different code sizes (between $L=9$ and $L=21$).}
    \label{multiple_rho}
\end{figure}

Fig.~\ref{multiple_rho} provides us with a more detailed version of Fig. \ref{cr_th_corrected}, where more values of $\rho$ are included. This confirms that the threshold follows a straight line from the right to the left when $\rho$ increases, as shown by the black dashed arrow of Fig. \ref{cr_th_corrected}.

\end{document}